\documentclass[aps,prd,groupedaddress]{revtex4}
\usepackage{array,hhline,dcolumn} 
\usepackage{rotating} 
\usepackage{hyperref}

\newcommand{\gtwid}{\mathrel{\raise.3ex\hbox{$>$\kern-.75em\lower1ex\hbox{$\sim$}}}}
\newcommand{\ltwid}{\mathrel{\raise.3ex\hbox{$<$\kern-.75em\lower1ex\hbox{$\sim$}}}}

\begin{document}
%

\title{MiniBooNE Data Releases}

\author{
        A.~A. Aguilar-Arevalo$^{14}$,
        B.~C.~Brown$^{5}$, 
        J.~M.~Conrad$^{13}$,
        R.~Dharmapalan$^{1,7}$, 
        A.~Diaz$^{13}$,
        Z.~Djurcic$^{2}$, 
        D.~A.~Finley$^{5}$, 
        R.~Ford$^{5}$,
        G.~T.~Garvey$^{10}$,
        S.~Gollapinni$^{10}$,
        A.~Hourlier$^{13}$, 
        E.-C.~Huang$^{10}$,
        N.~W.~Kamp$^{13}$, 
        G.~Karagiorgi$^{4}$, 
        T.~Katori$^{12}$,
        T.~Kobilarcik$^{5}$, 
        K.~Lin$^{4,10}$,
        W.~C.~Louis$^{10}$, 
        C.~Mariani$^{17}$, 
        W.~Marsh$^{5}$,
        G.~B.~Mills$^{10,\dagger}$,
        J.~Mirabal-Martinez$^{10}$, 
        C.~D.~Moore$^{5}$, 
        R.~H.~Nelson$^{3,\star}$, 
        J.~Nowak$^{9}$,
        I.~Parmaksiz$^{16}$, 
        Z.~Pavlovic$^{5}$, 
        H.~Ray$^{6}$, 
        B.~P.~Roe$^{15}$,
        A.~D.~Russell$^{5}$,
	    A.~Schneider$^{13}$,
        M.~H.~Shaevitz$^{4}$,
    	H.~Siegel$^{4}$,
        J.~Spitz$^{15}$, 
        I.~Stancu$^{1}$,
    	R.~Tayloe$^{8}$,
        R.~T.~Thornton$^{10}$, 
        M.~Tzanov$^{3,11}$,
        R.~G.~Van~de~Water$^{10}$,
        D.~H.~White$^{10,\dagger}$, 
        E.~D.~Zimmerman$^{3}$ \\
\smallskip
(The MiniBooNE Collaboration)
\smallskip
}
\smallskip
\smallskip
\affiliation{
$^1$University of Alabama; Tuscaloosa, AL 35487, USA \\
$^2$Argonne National Laboratory; Argonne, IL 60439, USA \\
$^3$University of Colorado; Boulder, CO 80309, USA \\
$^4$Columbia University; New York, NY 10027, USA \\
$^5$Fermi National Accelerator Laboratory; Batavia, IL 60510, USA \\
$^6$University of Florida; Gainesville, FL 32611, USA \\
$^7$University of Hawaii, Manoa; Honolulu, HI 96822, USA \\
$^8$Indiana University; Bloomington, IN 47405, USA \\
$^9$Lancaster University; Lancaster LA1 4YB, UK \\
$^{10}$Los Alamos National Laboratory; Los Alamos, NM 87545, USA \\
$^{11}$Louisiana State University; Baton Rouge, LA 70803, USA \\
$^{12}$King's College London; London WC2R 2LS, UK \\
$^{13}$Massachusetts Institute of Technology; Cambridge, MA 02139, USA \\
$^{14}$Instituto de Ciencias Nucleares; Universidad Nacional Aut\'onoma de M\'exico; CDMX 04510, M\'exico \\
$^{15}$University of Michigan; Ann Arbor, MI 48109, USA \\
$^{16}$University of Texas at Arlington, Arlington, TX 76019 \\
$^{17}$Center for Neutrino Physics; Virginia Tech; Blacksburg, VA 24061, USA \\
$^\star$Now at The Aerospace Corporation, Los Angeles, CA 90009, USA \\
$^\dagger$Deceased \\
}

\date{\today}

\begin{abstract}
The MiniBooNE experiment has provided data releases for most publications. Occasionally it is necessary to move data release pages. This document provides a single point of reference that will be updated by the collaboration to point to the present location of the MiniBooNE data releases.
\end{abstract}
\maketitle

The MiniBooNE experiment has provided data releases for most publications. From 2007-2021 these data releases were located on a website at Fermi National Accelerator Laboratory.  However, due to recent changes in policy at Fermilab, those data release web-pages are no longer available except to those with Fermilab services accounts.  

To enable continuous access, the data releases have been copied in the same form to a public site. The purpose of this note is to provide a citable arXiv entry pointing to the present location of the MiniBooNE data releases. This note will be updated as releases are moved or more releases are added. Many will be moved to \href{https://www.hepdata.net}{https://www.hepdata.net}

The releases listed below,  representing the complete set as of October 2021, are located at:   \href{https://rtayloe.pages.iu.edu/MB/data-releases/}{https://rtayloe.pages.iu.edu/MB/data-releases/}

\begin{itemize}

\item Data Released with A.A. Aguilar-Arevalo et al., ``Updated MiniBooNE Neutrino Oscillation Results with Increased Data and New Background Studies", arXiv:2006.16883 [hep-ex], Phys. Rev. D 103, 052002 (2021).

\item Data Released with A.A. Aguilar-Arevalo et al., ``Dark Matter Search in Nucleon, Pion, and Electron Channels from a Proton Beam Dump with MiniBooNE", arXiv:1807.06137 [hep-ex], Phys. Rev. D 98, 112004 (2018).

\item Data Released with A.A. Aguilar-Arevalo et al., ``Observation of a Significant Excess of Electron-Like Events in the MiniBooNE Short-Baseline Neutrino Experiment", arXiv:1805.12028 [hep-ex], Phys. Rev. Lett. 121, 221801 (2018).

\item Data Released with A.A. Aguilar-Arevalo et al., ``First Measurement of Monoenergetic Muon Neutrino Charged Current Interactions", arXiv:1801.03848 [hep-ex], Phys. Rev. Lett. 120, 141802 (2018).

\item Data Released with A.A. Aguilar-Arevalo et al., ``First Measurement of the Muon Antineutrino Double-Differential Charged-Current Quasielastic Cross section", arXiv:1301.7067 [hep-ex], Phys. Rev. D88, 032001 (2013).

\item Data Released with A.A. Aguilar-Arevalo et al., ``Improved Search for $\bar \nu_\mu \rightarrow \bar \nu_e$ Oscillations in the MiniBooNE Experiment", arXiv:1303.2588, Phys. Rev. Lett. 110, 161801 (2013).

\item Data Released with A.A. Aguilar-Arevalo et al., ``Event Excess in the MiniBooNE Search for numubar to nuebar Oscillations", arXiv:1007.1150 [hep-ex], Phys. Rev. Lett. 105, 181801 (2010).

\item Data Released with A.A. Aguilar-Arevalo et al., ``Measurement of Neutrino-Induced Charged Current-Charged Pion Production Cross Sections on Mineral Oil at Enu $\sim$ 1 GeV", arXiv:1011.3572 [hep-ex], Phys. Rev. D 83, 052007 (2011).

\item Data Released with A.A. Aguilar-Arevalo et al., ``Measurement of Muon Neutrino-Induced Charged-Current Neutral Pion Production Cross Sections on Mineral Oil at Enu = 0.5-2.0 GeV", arXiv:1010.3264 [hep-ex], Phys. Rev. D 83, 052009 (2011).

\item Data Released with A.A. Aguilar-Arevalo et al., ``Measurement of the Neutrino Neutral-Current Elastic Differential Cross Section", arXiv:1007.4730 [hep-ex], Phys. Rev. D 82, 092005 (2010).

\item Data Released with A.A. Aguilar-Arevalo et al., ``First Measurement of the Muon Neutrino Charged-Current Quasielastic Double Differential Cross section", arXiv:1002.2680 [hep-ex], Phys. Rev. D81, 092005 (2010).

\item Data Released with A.A. Aguilar-Arevalo et al., ``Measurement of $\nu_\mu$  and$\bar\nu_\mu$  Induced Neutral-Current Single $\pi^0$ Production Cross Sections on Mineral Oil at $E_\nu\sim$O(1 GeV)", arXiv:0911.2063 [hep-ex], Phys. Rev. D81, 013005 (2010).

\item Data Released with A.A. Aguilar-Arevalo et al., ``A Search for Electron Anti-Neutrino Appearance at  $\Delta m^2\sim 1$ eV$^2$  Scale", arXiv:0904.1958 [hep-ex], Phys. Rev. Lett. 103, 111801 (2009).

\item Data Released with A.A. Aguilar-Arevalo et al., ``A Search for Muon Neutrino and Anti-Neutrino Disappearance in MiniBooNE", arXiv:0903.2465 [hep-ex], Phys. Rev. Lett. 103, 061802 (2009).

\item Data Released with A.A. Aguilar-Arevalo et al., ``Unexplained Excess of Electron-Like Events From a 1 GeV Neutrino Beam", arXiv:0812.2243 [hep-ex], Phys. Rev. Lett. 102, 101802 (2009).

\item Data Released with A.A. Aguilar-Arevalo et al., ``The Neutrino Flux Prediction at MiniBooNE", arXiv:0806.1449 [hep-ex], Phys. Rev. D. 79, 072002 (2009).

\item Data Released with A.A. Aguilar-Arevalo et al., ``A Search for Electron Neutrino Appearance at the $\Delta m^2\sim 1$ eV$^2$ Scale", arXiv:0704.1500 [hep-ex], Phys. Rev. Lett. 98, 231801 (2007). 
  
\end{itemize}

\end{document}